\documentclass[preprint,prb,showpacs,preprintnumbers,amsmath,amssymb]{revtex4}

\usepackage{graphicx}
\usepackage{dcolumn}
\usepackage{bm}
\usepackage[usenames]{color}

\begin{document}


\title{Influence of Anharmonic Effects on the Zero-point Vacancy Concentration in Solid $^4$He}

\author{R. Pessoa}
\email{rpessoa@ifi.unicamp.br}
\author{M. de Koning}
\email{dekoning@ifi.unicamp.br}
\author{S. A. Vitiello}
\email{vitiello@ifi.unicamp.br}
\affiliation{Instituto de F\'{\i}sica Gleb Wataghin, Caixa Postal 6165 \\
Universidade Estadual de Campinas - UNICAMP \\
13083-970, Campinas, SP, Brazil}

\date{\today}

\begin{abstract}
 We conduct a theoretical study in which we determine the zero-point vacancy concentration in solid $^{4}$He at $T=0 K$. To this end, we employ the quantum-classical isomorphism, by which the quantum-mechanical probability density function of a system composed of bosons at $T=0K$ can be interpreted in terms of a Boltzmann factor of a classical system at finite temperature. By using this classical isomorph we apply the methods of classical statistical mechanics to compute the vacancy formation free energy and the vacancy concentration in the associated quantum system at $T=0$. In this context, we focus specifically on the role of anharmonic effects that are expected to be non-negligible due to the significant zero-point motion. For this purpose, we compute the formation free energies using both the harmonic approximation (HA) as well as reversible-work (RW) method, in which all anharmonic effects are taken into account. The results suggest that anharmonic effects indeed play a significant role, lowering the classical formation free energy by $\sim 25\%$ and increasing the zero-point vacancy concentration by more than an order of magnitude compared to the HA.
\end{abstract}


\maketitle

\section{Introduction}
\label{intro}

The unusual properties of the condensed phases of Helium have turned this substance into one of the most studied in the literature. An example is the fact that Helium does not exhibit a solid phase at ambient pressure, not even at temperatures tending to absolute zero. This is due to the weak interactions between the Helium atoms and the large zero-point motion. Accordingly, only the application of a significant external pressure, of the order of 25 atm, allows the formation of a stable solid phase. The high-pressure solid phase of $^{4}$He has recently become the subject of intensive investigation after the experimental observations of nonclassical rotational inertia (NCRI)~\cite{kim04,kim04b,kim06,cha08} and unexpected mechanical behavior.~\cite{DayNature07}

Although $^{4}$He's liquid phase and its display of the phenomenon of superfluidity are comparatively well understood, the comprehension of its solid counterpart involves more subtle issues that are not so easily grasped (for a recent review see Ref.~\onlinecite{whi06}). This is reflected, for instance, in the fact that the origin of the observed NCRI continues to be a subject of debate. A possible interpretation of this phenomenon is the existence of a supersolid phase, which presents the long-range order and spontaneous translational and rotational symmetry breaking characteristic of a solid but displaying superfluid-like behavior. The pioneering theoretical proposals of a $^{4}$He supersolid phase~\cite{and69,che70} predict that vacancies are required for quantum crystals to exhibit superfluid behavior. More recently, despite of some controversy,~\cite{bon06,mah06} there is evidence~\cite{gal06} that vacancies can indeed exist at zero temperature and that they may play a role in the observed NCRI.

In view of the possible role of lattice vacancies in the unusual behavior of solid $^{4}$He, it is of interest to estimate their zero-temperature equilibrium concentrations. Given that experimental estimates are not yet available, we need to resort to purely theoretical methods. Hodgdon and Stillinger~\cite{hod95} were the first to conduct such an effort, taking advantage of the well-known isomorphism between a quantum-mechanical system composed of Bosons at zero temperature and a classical system at finite temperature. This isomorphism is a consequence of the fact that the wave function describing the ground state of a system of identical Bosons is nodeless.~\cite{cam77} Accordingly, the associated probability density function can be formally written in terms of a classical Boltzmann factor $\exp(-U/k_{B}T)$, where $U$ is the
classical potential energy function, $T$ is the absolute temperature, $k_{B}$ is the Boltzmann constant and $k_{B}T$ is to be taken equal to $1$. By exploiting this isomorphism it is possible to apply the finite-temperature methods of classical statistical mechanics to estimate the properties of a quantum-mechanical Boson system at zero temperature. Specifically, the calculation of the zero-point equilibrium vacancy concentration in solid $^{4}$He at zero temperature becomes equivalent to the determination of the thermal equilibrium vacancy concentration of the associated classical solid at a finite temperature. The latter can be computed by determining the finite-temperature vacancy-formation free energy.

Hodgdon and Stillinger~\cite{hod95} utilized the standard classical harmonic approximation (HA)~\cite{Maradudin,Foiles1994} to determine the vacancy formation free energy and corresponding vacancy concentration. In classical solids, however, the HA is known to be problematic in the presence of defects,~\cite{Foiles1994,deKoning2004,deDebiaggi2006} which, due to the local breaking of lattice symmetries, enhance the role of anharmonic effects. In this context, given the significant zero-point motion in quantum solids, such anharmonic effects are also expected to be relevant in the determination of the zero-temperature equilibrium concentration of lattice vacancies in solid $^{4}$He. In this light, the purpose of the present paper is to evaluate the influence of anharmonic effects on the zero-point vacancy concentration in solid $^{4}$He. To this end, following the approach of Hodgdon and Stillinger,~\cite{hod95} we apply the aforementioned isomorphism using an approximate description of solid $^{4}$He based on a trial function of the Jastrow-form.~\cite{mcm65,cep79} However, in addition to applying the HA to compute finite-temperature vacancy-formation free energy in the associated classical solid, we also utilize an exact method in which all anharmonic effects are explicitly included.

The remainder of the paper has been organized as follows. Section~\ref{metho} discusses the details of the isomorphism between a quantum solid composed of Bosons in its ground state and a classical crystal. Furthermore, it gives a description of the computational methods used to determine the formation free energy and the corresponding thermal equilibrium concentration of vacancies in a classical crystal. Section \ref{details} describes the computational details of the calculations and in Section \ref{results} we discuss the obtained results. We conclude with a summary in Section~\ref{final}.

\section{Methodology}
\label{metho}
\subsection{The isomorphism between classical and quantum Bose systems}

The simplest trial wave function that is able to describe a system of $^{4}$He atoms is a pairwise function of the Jastrow form
\begin{equation}
 \Psi({\bf R}) = \exp\left[- \frac{1}{2} \sum_{i<j}^{N} u(r_{ij})\right] \; ,
\label{jastrow}
\end{equation}
where ${\bf R} \equiv \left\lbrace {\bf r}_{1},{\bf r}_{2}, \ldots, {\bf r}_{N} \right\rbrace $ represents the $N$ coordinates of the atoms of the system and $u(r)$ is the so-called pseudopotential.~\cite{mcm65,cep79} In all our calculations we utilize the particular form
\begin{equation}
u(r)=\left( \frac{b}{r}\right)^{6} ,
\label{pseudo}
\end{equation}
containing the single parameter $b$.~\cite{mcm65} Although pseudopotentials that give better variational energies for the solid phase do exist,~\cite{nos64,vit88} we have chosen this particular representation because it has been already used in HA calculations and for its computational simplicity. The methodology utilized to evaluate the influence of anharmonic effects, however, is not restricted to this particular form and can be applied to any of the more accurate representations proposed in the literature.

The isomorphism that exists between the probability density of finding a quantum system of identical Bosons in the ground-state as described by Eq.~(\ref{jastrow}) and a classical crystal of potential energy $U({\bf R})$ at temperature $T$ is easily realized by writing
\begin{equation}
\label{potential_energy}
 U({\bf R})=\sum_{i<j}u(r_{ij})\;\; ,
\end{equation}
such that
\begin{equation}
\label{Boltzmann}
 \vert \Psi \vert^{2}= \exp\left( - \frac{U({\bf R})}{k_{B}T}\right),
\end{equation}
with $k_B T=1$. Here, and in the remainder of the paper, the potential energy $U$ and all quantities that have the dimension of energy are measured in arbitrary energy units. 
 Eq.~(\ref{Boltzmann}) can be interpreted as the Boltzmann factor for a configuration ${\bf R}$ of a classical system described by the potential energy function Eq.~(\ref{potential_energy}). For the particular form of the pseudopotential used in the present paper, Eq.~(\ref{pseudo}), the classical system corresponds to a collection of purely repulsive soft spheres.

\subsection{Vacancy thermodynamics in classical systems}
\label{thermo}

The determination of the thermal equilibrium concentration of vacancies in a classical crystal requires the minimization of the free energy of a crystal containing $N$ atoms with respect to the total number of vacancies $n$.~\cite{Ashcroft}
Let $f$ be the formation free energy of a monovacancy in the crystal, defined as
\begin{equation}
 f=F(N-1)-\frac{N-1}{N}F(N),
\label{formation_f}
\end{equation}
where $F(N-1)$ and $F(N)$ are the Helmholtz free energies of, respectively, a crystal containing $N-1$ atoms and a single
vacancy, and a defect-free crystal containing $N$ atoms. Accordingly, the free energy of a system consisting of $N$ atoms and $n$ vacancies can be written as
\begin{equation}
 F(N+n)= F(N)+nf-k_{B}T \ln\left[ \frac{(N+n)!}{n!N!}\right] \;.
\label{fNn}
\end{equation}
Note that when the number of vacancies is zero in the above expression, $F(N+n)$ reduces to the free energy of the defect-free system. The last term in Eq.~(\ref{fNn}) is just $TS_{\rm conf}$ where $S_{\rm conf}$ represents the configurational entropy associated with the number of ways in which the $n$ vacancies can be distributed among the lattice sites. In thermodynamic equilibrium the free energy is a minimum and the number of vacancies may be determined by minimizing Eq.~(\ref{fNn}) with respect to $n$. The resulting thermal equilibrium vacancy concentration $c$ is given by
\begin{equation}
c\equiv\frac{n}{N}=\exp\left[-\frac{f}{k_{B}T} \right]  .
\label{concentration}
\end{equation}

\subsection{Free-energy calculations}

\subsubsection{Reversible work}

In order to determine the vacancy concentration by means of Eq.~(\ref{concentration}), we need to compute the vacancy-formation free energy $f$. Given that the free energy is a thermal quantity that cannot be expressed in terms of an ensemble average,~\cite{all89,fre02} it cannot be computed directly using any Monte Carlo or Molecular Dynamics sampling method. As a result, free energies are usually determined using indirect strategies, in which free-energy {\em differences} between two systems are computed by evaluating the work associated with a reversible process that connects these two systems.~\cite{fre02} This approach has shown to be very effective for the computation of defect thermal equilibrium concentrations in classical solids.~\cite{fre02,deKoning2004,deDebiaggi2006} Accordingly, by virtue of the discussed quantum-classical isomorphism, it should also prove useful for the evaluation of zero-point defect concentrations in Boson quantum solids.

The main idea of the reversible work (RW) method is to take a reference system, for which
the free energy is known, and connect it to the system of interest by means of a coupling
parameter $\lambda$. A typical functional form of this coupling is given by the Hamiltonian~\cite{fre02}
\begin{equation}
 H(\lambda)=\lambda H_{0}+(1-\lambda)H_{\rm ref} ,
\label{coupled}
\end{equation}
where $H_{0}$ and $H_{\rm ref}$ represent the Hamiltonians of the system of interest and of reference, respectively. Note that this form allows a continuous switching between $H_{\rm ref}$ and $H_{0}$ by varying the parameter $\lambda$, varied between $0$ and $1$. The free-energy difference between the system of interest and the reference is then given by the {\em reversible} work $W_{\rm rev}$~\cite{fre02}
\begin{equation}
 W_{\rm rev} \equiv F_0-F_{\rm ref}=\int_0^1 d\lambda \left< \frac{\partial H}{\partial \lambda} \right>,
\label{reversible work}
\end{equation}
where the brackets indicate an equilibrium average in the relevant statistical ensemble (\textit{i.e.}, canonical, isobaric-isothermal, \textit{etc.}). The integration represents the calculation of the total work done by the generalized force $\partial H/\partial \lambda$. Since the integration involves equilibrium averages of the system at all times, it reflects a reversible process.

In principle, the numerical evaluation of the work integral Eq.~(\ref{reversible work}) requires a series of independent equilibrium simulations, each carried out at a different value of the coupling parameter $\lambda$ between 0 and 1.~\cite{fre02} In practice, however, it has shown beneficial to estimate the work integral along a single, nonequilibrium simulation during which the value of $\lambda$ changes dynamically. In this fashion, the reversible work $W_{\rm rev}$ is estimated in terms of the {\em irreversible} work estimator
\begin{equation}
 W_{\rm irr} = \int_{0}^{t_{\rm sim}}dt' \left[ \frac{d\lambda}{dt}\right]_{t'} \left[ \frac{\partial H(\{ \textbf{r}_{i}\},\lambda)}{\partial \lambda}\right]_{\lambda(t')}  .
\label{work_irr}
\end{equation}
where $t'$ represents the time coordinate that describes the dynamical evolution of the coupling parameter $\lambda(t)$, and $t_{\rm sim}$ is the total duration of the switching process.

Given that the dynamical process above is intrinsically irreversible, the irreversible work estimator~(\ref{work_irr}) is {\em biased}, subject to a positive systematic error that is associated with the dissipative entropy production inherent to nonequilibrium processes.~\cite{deKoning1997,deKoning2000,deKoning2005} As a result, it will represent an upper bound to the value of the reversible work $W_{\rm rev}$ and, consequently, the free-energy difference $F_0-F_{\rm ref}$. Fortunately, however, the systematic error can be readily eliminated, as long as the nonequilibrium process remains within the regime of linear response. In this regime, the systematic error is independent of the direction of the switching process, i.e., the entropy production is equal for the forward ($\lambda=0\rightarrow 1$) and backward ($\lambda=1\rightarrow 0$), processes.~\cite{deKoning2000,deKoning2005} In this fashion, we can obtain an unbiased estimate for $W_{\rm rev}$ according to
\begin{equation}
 W_{\rm rev} = \frac{1}{2}\left[ W_{\rm irr}(\lambda=0 \rightarrow 1) - W_{\rm irr}(\lambda= 1 \rightarrow 0) \right] ,
\label{w}
\end{equation}
which is subject to statistical errors only.~\cite{deKoning2000,deKoning2005}

The initial step toward computing the formation free energy of a vacancy, Eq.~(\ref{formation_f}), utilizing the RW approach involves the calculation of the free energy of a defect-free crystal containing $N$ atoms of $^4$He, $F(N)$. The strategy will be to use the classical Einstein crystal, of which the free energy $F_{\rm Einst}$ is known analytically, as a reference system.~\cite{fre02} In this manner, using the classical potential energy expression of Eq.~(\ref{potential_energy}), the particular form of the coupled Hamiltonian (\ref{coupled}) becomes
\begin{equation}
 H \left( {\bf R} ; \lambda\right)  = \lambda \left[\sum_{i<j}^{N}
 \left( \frac{b}{r_{ij}} \right)^{6}\right] +
(1-\lambda) \left[ {\cal U}({\bf R}^{(0)}) + \frac{1}{2} \sum_{i=1}^{N} \kappa_{i}
 \left( {\bf r}_{i} - {\bf r}_{i} ^ {(0)} \right)^{2}  \right]  ,
\label{H_eins}
\end{equation}
where ${\bf R}^{(0)}$ represents the equilibrium lattice positions of $N$ atoms, ${\cal U}({\bf R}^{(0)})$
is the static contribution to the potential energy and $\kappa_{i}$ is a spring constant.

Since $W_{\rm rev}$ is the difference between the Helmholtz free energy of the system of interest
and the reference system, the free energy of the defect-free crystal is given by
\begin{equation}
 F(N) = W_{\rm rev} + F_{\rm Einst} \; .
 \label{FN}
\end{equation}
It is convenient to fix the center of mass of the system during the switching process.~\cite{fre02} In the presence of
this constraint, the free energy of the Einstein crystal is given by
\begin{equation}
 F_{\rm Einst} = {\cal U}({\bf R}^{(0)}) - \frac{3}{2} k_{B}T\sum_{i=1}^{N}
\ln\left[ \frac{2 \pi k_{B}T}{\kappa_i}\right] -
\frac{3}{2} k_{B}T \ln\left[ \frac{\kappa_i}{8 \pi^{2} (k_{B}T)^2}\right] .
\end{equation}
The value of the spring constants $\kappa_i$ is adjusted such that the mean-square displacement of the harmonic oscillator is approximately equal that of the particles in the system of interest.~\cite{fre02}

In the second step we estimate the free energy $F(N-1)$ of the system
containing a single vacancy. In order to apply the RW method for this purpose, we use a
strategy that is similar to the one employed to compute $F(N)$. In this case, the parameter $\lambda$ controls the strength of the interaction between a single particle and all the others. The coupled Hamiltonian for this process is given by
\begin{equation}
 H \left( \{ { \bf r}\} ; \lambda\right)  = \mathop{\sum_{{i<j}}^{N}}_{(i\neq k)}
 \left( \frac{b}{r_{ij}} \right)^{6} +
\lambda \mathop{\sum^{N}_{i}}_{(i\neq k)}
 \left( \frac{b}{r_{ik}} \right)^{6} +
(1-\lambda) \frac{1}{2} \kappa
 \left( {\bf r}_{k} - {\bf r}_{k} ^ {(0)} \right)^{2} ,
\label{H_ws}
\end{equation}
where the index $k$ stands for the particle that is turned on and off as $\lambda$ is varied between 0 and 1. The final
term describes a single harmonic oscillator, which is the final state of particle $k$ when it has been totally decoupled from
the remainder of the system. The coupling to the harmonic oscillator is implemented to prevent the decoupled particle $k$ from drifting
freely through the system.~\cite{deKoning2004} As before, the center of mass is held fixed during the switching simulations.
Furthermore, it is convenient~\cite{hod95} to maintain the particle density $\rho$ of the system constant. This is
done by varying the size $L$ of the simulation box as the value of $\lambda$ is changed. Specifically, we set
\begin{equation}
L(\lambda)=\sqrt[3]{\frac{N-1+\lambda}{\rho}},
\label{dens_cte}
\end{equation}
such that the density of {\em interacting} particles is equal to $\rho$ both for $\lambda=0$ and $\lambda=1$.

A final concern involves the applicability of the approach at elevated
temperatures, where vacancy diffusion may interfere with the RW
process. In this situation, one of the twelve nearest-neighbor atoms of the decoupled particle may move
into the vacant lattice site by means of a diffusion event, which leads to singularities in the driving force.~\cite{deKoning2004}
An adequate solution to this problem is to constrain the motion of the neighboring atoms such that they are forced to remain
within their respective Wigner-Seitz primitive cells.~\cite{deKoning2004}

From the coupled Hamiltonian in Eq.~(\ref{H_ws}) we determine the
driving force $\partial H/\partial \lambda$ and compute the reversible work $W_{\rm rev}'$ required to turn off the interactions
between particle $k$ and the remainder of the system and turn it into an independent harmonic oscillator. As in the previous case,
the switching process is carried out in both directions to eliminate the systematic error. Accordingly,
$F(N-1)$ is obtained through
\begin{equation}
\label{FN-1}
 F(N-1)= W_{\rm rev}'+F(N)-F_{\rm osc} \; ,
\end{equation}
where $F_{\rm osc}$ is the free energy of a single harmonic oscillator, given by~\cite{fre02}
\begin{equation}
 F_{\rm osc}=-\frac{3}{2} k_{B}T \ln \left[ \frac{2 \pi k_{B}T}{\kappa}\right] .
\end{equation}

Finally, combining the results of Eqs.~(\ref{FN}), and (\ref{FN-1}) we compute the vacancy formation free energy
through Eq.~(\ref{formation_f}). The corresponding zero-point vacancy concentration is then given by Eq.~(\ref{concentration}).

\subsubsection{Harmonic approximation}

The harmonic approximation~\cite{Maradudin} is a method that allows one to estimate the vacancy-formation free energy without sampling configurations from a statistical ensemble. Hodgdon and Stillinger~\cite{hod95} used this approach to estimate the concentration of point defects in solid $^{4}$He described by a Jastrow wave function. The central idea of this approximation is to write the free energy as the sum of static potential energy plus a finite-temperature part involving the vibrational normal mode frequencies $\omega_{i}$, where $i=1,\ldots,3N-3$.

In this manner,~\cite{hod95} the vacancy formation $f$ is given by
\begin{eqnarray}
 f &=& {\cal U}(N-1) - \left( \frac{N-1}{N} \right) {\cal U}({\bf R}^{(0)}) \nonumber \\
&+& k_{B}T\sum_{i=1}^{3N-6} ln\left[ \frac{\omega_{i}(N-1)}{k_{B}T}\right] - \left( \frac{N-2}{N-1} \right) k_{B}T \sum_{i=}^{3N-3} ln \left[ \frac{\omega_{i}(N)}{k_{B}T}\right] \;,
\end{eqnarray}
where ${\cal U}(N-1)$ represent the static potential energies of the crystal containing a single vacancy. Similarly, the last two terms represent the vibrational parts of the crystals with and without vacancy, respectively. In practice the vibrational frequencies are obtained by diagonalizing the Hessian matrix of the potential-energy functions.~\cite{hod95}

\section{Computational Details}
\label{details}

All calculations were carried out using fcc cubic computational cells containing numbers of particles $N$ varying between $108$ and $1372$. To eliminate surface effects, we apply standard periodic boundary conditions and invoke the minimum-image convention.~\cite{fre02} In order to be consistent with a continuous and differentiable wave function~\cite{vit99} we force the pseudopotential to be equal to zero for interatomic distances $r=L/2$. A convenient way to do is to slightly modifying the pseudopotential $u(r)$ of Eq.~(\ref{pseudo}) according to
\begin{equation}
u(r)\rightarrow \tilde{u}(r) = u(r)+u(L-r)-2u(L/2)\;\; .
\label{util}
\end{equation}
In most of our calculations we have chosen $b=1.72\sigma$, with $\sigma=2.556 \AA{}$, for which the classical system crystallizes into a $fcc$ structure, as determined by Prestipino \textit{et al.}.~\cite{pre05} In practice, the system remains a metastable $fcc$ solid for smaller values of $b$, and we have done a few additional calculations for smaller $b$-values. Hodgdon and Stillinger,~\cite{hod95} for instance, chose the parameter $b=1.67\sigma$ for which the $0$K melting density of $^4$He equals the melting density of the classical crystal as obtained by Hoover \textit{et al.}.~\cite{hoo71} We find, however, that this value of $b$ does not describe a metastable fcc solid and leads to a liquid phase.\cite{Note}

We apply the Metropolis algorithm to sample the configurations from $\vert\Psi(R)\vert^{2}$.
In the RW switching simulations, the switching parameter is varied linearly according to $\lambda=t/t_{\rm sim}$ (forward process) or $\lambda=1-t/t_{\rm sim}$ (backward process). Here, $t$ is an index of the MC sweep in the process and $t_{\rm sim}$ represents the total number of sweeps of the switching process. It is important to remember that one must always start an RW switching process from an equilibrated
system. Accordingly, every RW switching run, for both switching directions, was preceded by an equilibration run of at least $5\times 10^3 MC$ sweeps before computing the work Eq.~(\ref{work_irr}). In order to guarantee that the RW results obtained from nonequilibrium processes are bias-free, we compute the irreversible work estimators Eq.~(\ref{work_irr}) for both switching directions as a function of the total switching duration $t_{\rm sim}$ and monitor the convergence of Eq.~(\ref{w}). For each value of $t_{\rm sim}$ we determine averages of the irreversible work estimator over 30 switching processes.

\section{Results and Discussion}
\label{results}

\begin{figure} [c]
\includegraphics[angle=-90,width=\linewidth]{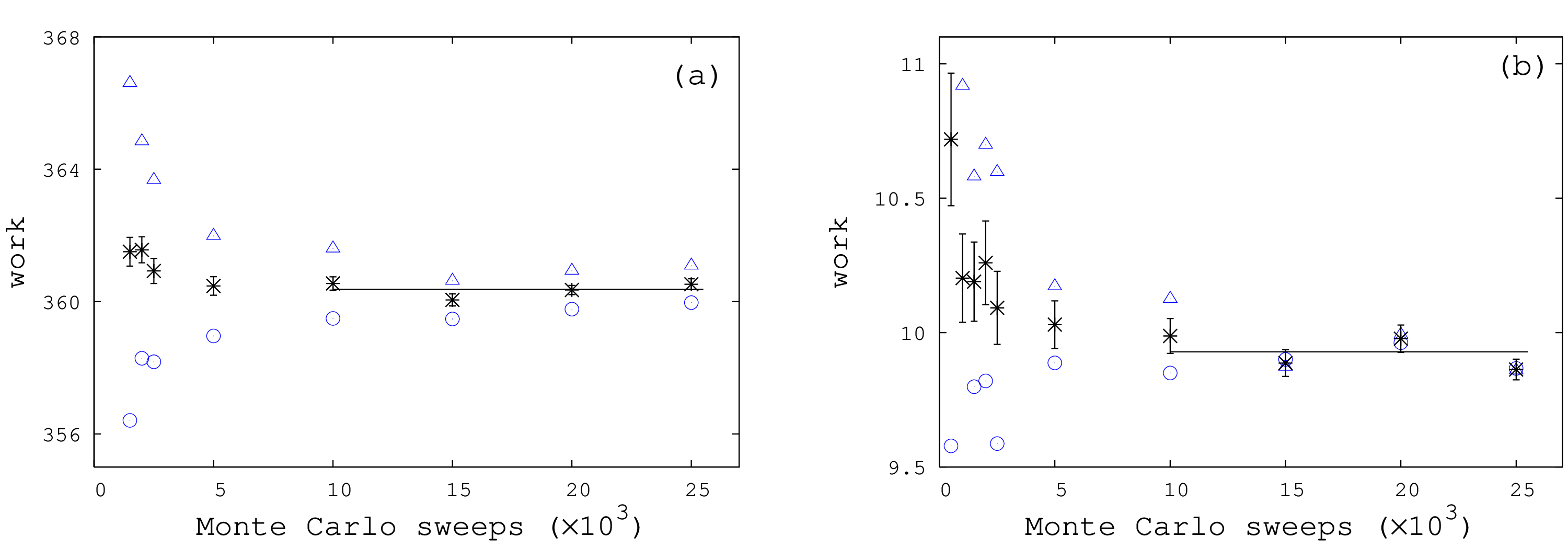}%
\caption{\label{work} Irreversible work calculations associated with the process of transforming the interacting, defect-free solid into an Einstein crystal (Panel $a$) and the process of introducing a monovacancy into the defect-free solid (Panel $b$) as a function of the process duration $t_{\rm sim}$ measured in MC sweeps. Results are displayed for the forward processes ({\color{blue} $\triangle$}), backward processes ({\color{blue}$\bigcirc$}), as well as the corresponding unbiased estimator of Eq.~(\ref{w}) ($\ast$). Each point was obtained as the average over thirty independent simulations carried out $N=500$ particles. The solid line is an average of the last four values of the unbiased work estimator, for which the processes have reached the linear-response regime.}
\end{figure}

Fig.~\ref{work} shows the results of the RW calculations for the defect-free crystal (panel a) as well as for the reversible vacancy insertion process (panel b) obtained for the cell containing 500 atoms, with $k_B T=1$. The graphs display the average values of the forward and backward irreversible work estimators, as well as the unbiased estimator Eq.~(\ref{w}), as a function of the total process duration $t_{\rm sim}$. The results show that $t_{\rm sim} \approx 10^{4}$ sweeps is sufficiently long for the linear response regime to be reached in both cases, for which the unbiased estimator reaches convergence. The horizontal lines on the graphs are averages of the last four values of the reversible work in that region. Based on these results, all subsequent RW switching simulations based on different numbers of particles and temperatures were carried out with $t_{\rm sim}=1.5\times10^{4}$ MC sweeps.

\begin{figure} [c]
\includegraphics[angle=-90,width=12.0cm]{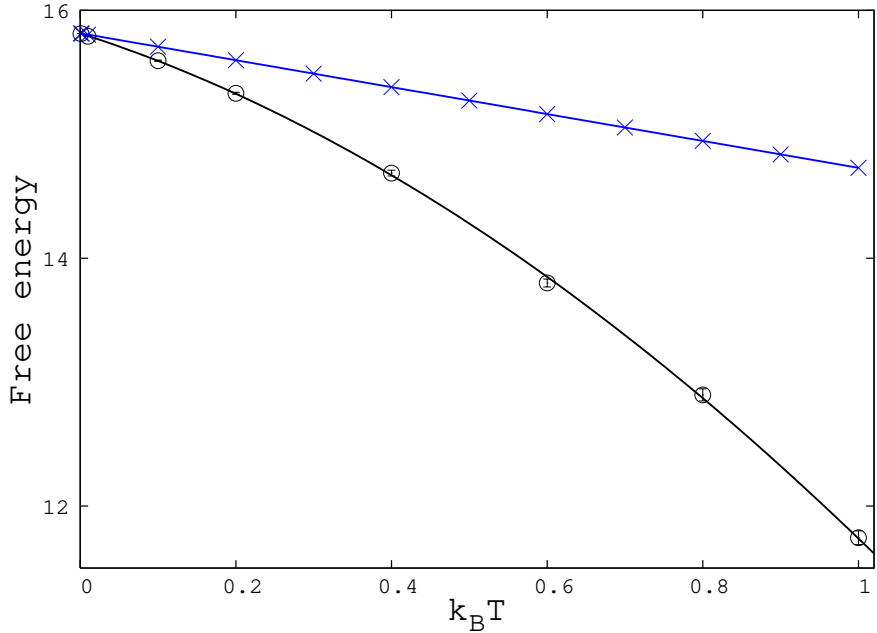}%
\caption{\label{f_QHA_sw} Vacancy-formation free energy as a function of temperature. Results were obtained for $b = 1.72\sigma$ and $N=500$. HA ({\color{blue} $\times$}) and RW ($\bigcirc$) results. The lines are guides to the eyes.}
\end{figure}

The results depicted in Fig.~\ref{f_QHA_sw} show a comparison between the vacancy-formation free energy as a function of temperature as obtained using the RW and HA methods, for a cell containing 500 particles. The data clearly demonstrate the significance of anharmonicity. As expected, for low temperatures, anharmonic effects are small and  both the HA and the RW method converge to the same formation free-energy value for zero temperature. As temperature increases, however, anharmonicity becomes increasingly important and the deviation between the RW and HA results becomes significant. Specifically, for $k_{B}T=1$, for which the isomorphism between the classical and
quantum description of the system is valid, anharmonic effects are responsible for 
a reduction of about $25\%$ in the value of the vacancy-formation free energy compared to the HA result. The anharmonic effects are associated with the significant zero-point motion in the quantum system at $T=0~K$. The extent of this motion can be approximately estimate in terms of the formation entropy of the classical system, $s=-\frac{\partial f}{\partial T}$, at $k_B T=1$, which can be computed from the slope of the formation free-energy curves. The resulting values give  $s=1.1 k_B$ and $s=6.0 k_B$ for the HA and RW results, respectively. The difference of almost a factor of six
is another clear indication of the importance of anharmonicity in the system.

\begin{figure} [c]
\includegraphics[angle=-90,width=12.0cm]{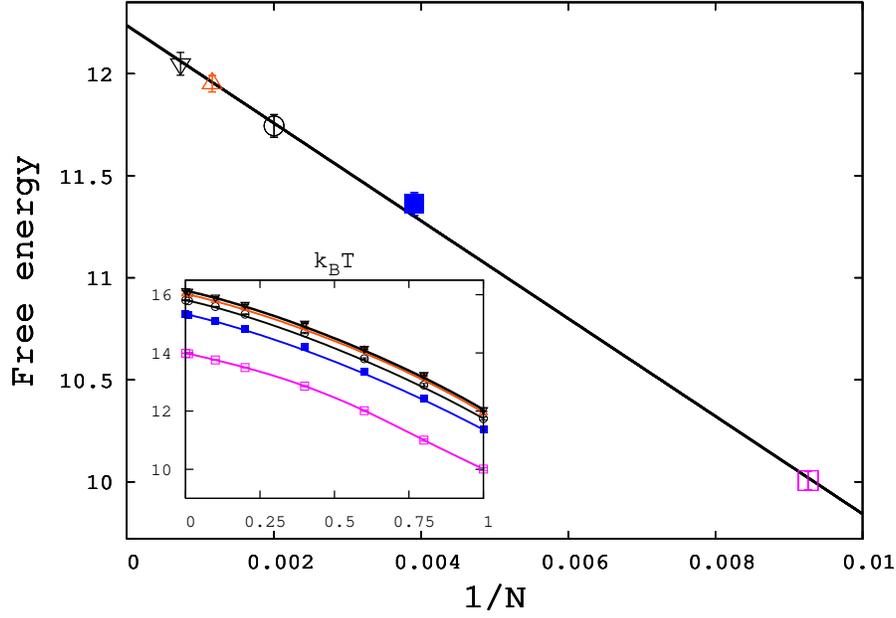}%
\caption{\label{free_size} Vacancy-formation free energy as a function of system size at $k_{B}T=1$. The straight was obtained through a least-square linear fit.
Results obtained for $b=1.72\sigma$ and $N=108$ ({\color{magenta} $\square$}), $N=256$ ({\color{blue} $\blacksquare$}), $N=500$ ({\color{black} $\bigcirc$}), $N=864$ ({\color{Orange} $\bigtriangleup$}) and $N=1372$ ($\bigtriangledown$). Inset shows the formation free energy as a function of $k_{B}T$ for the mentioned system sizes.}
\end{figure}

To assess the influence of finite-size effects, we computed the RW vacancy-formation free energy at $k_B T=1$ for different cell sizes, ranging from $N=108$ to $N=1372$ atoms. The results are shown in Fig.~\ref{free_size}, that plots the formation free energy as a function of  $1/N$. The inset shows the full formation free-energy versus temperature curves for different cell sizes. The results allow us to extrapolate the vacancy-formation free energy to $1/N \rightarrow 0$, which gives $f=12.24$ for infinite cell size.

\begin{figure} [c]
\includegraphics[angle=-90,width=12.0cm]{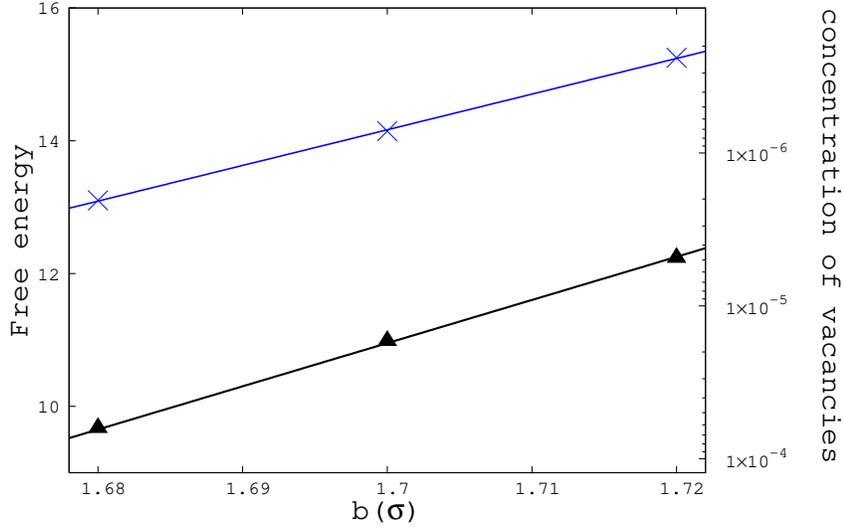}%
\caption{\label{free} Extrapolated vacancy-formation free energy as a function of $b$ at $k_{B}T=1.0$. Results obtained using the RW method ($\blacktriangle$) and the HA ({\color{blue} $\times$}). Lines are plotted to guide the eyes.}
\end{figure}

The determination of the extrapolated vacancy-formation free energy at $k_B T=1$ now allows us to compute the zero-point vacancy concentration for the quantum system described by Eqs.~(\ref{jastrow}) and (\ref{pseudo}) by using Eq.~(\ref{concentration}). Aside from the RW data, we have also used extrapolated HA formation free-energy data (not shown in Fig.~\ref{free_size}) for the calculation of the equilibrium concentration. In addition, we have also considered the dependence of the vacancy concentration on the value of the parameter $b$. The results are shown in Fig.~\ref{free}, which 
plots the extrapolated vacancy-formation free energy as well as the associated vacancy concentration as a function of the parameter $b$, using both the RW and HA methods. The effects of anharmonicity are clearly visible, reducing the formation free energy by $\sim 20-25\%$ and increasing the vacancy concentration by more than an order of magnitude. The results are also seen to be quite sensitive to the value of the parameter $b$, which is to be expected given that variations in $b$ can be interpreted in terms of variations in the temperature of the associated classical isomorph.    

\begin{figure} [c]
\includegraphics[angle=-90,width=13.0cm]{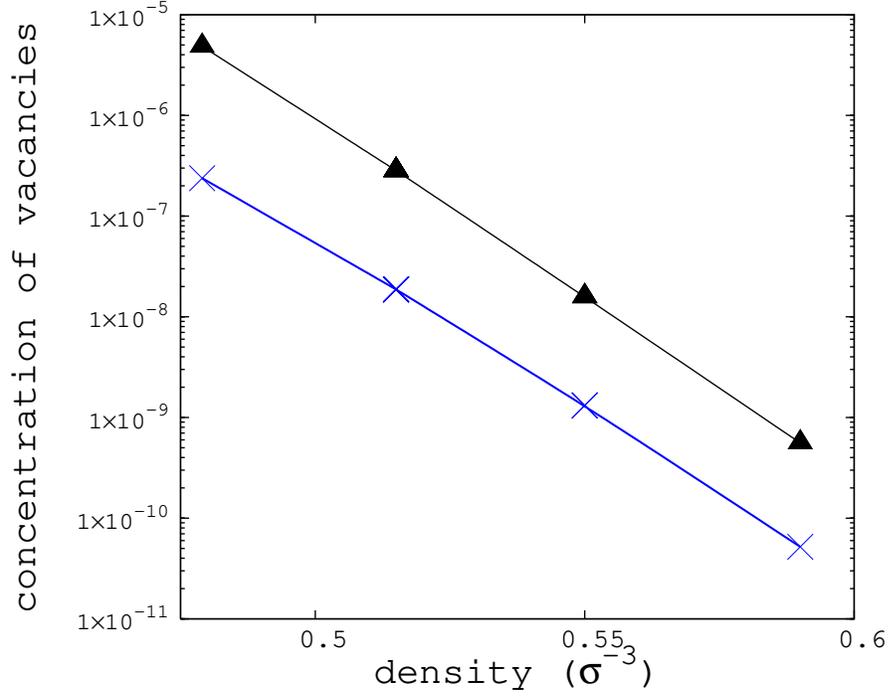}%
\caption{\label{c_dens} Concentration of vacancies as a function of density for $b=1.72\sigma$. Results obtained using RW method ($\blacktriangle$) and HA({\color{blue} $\times$}). Lines are plotted to guide the eyes.}
\end{figure}

We have also determined the vacancy concentration for densities above the $^4 He$ melting density. For this purpose we have computed the extrapolated vacancy-formation free energies as a  function of the density of the classical soft-sphere isomorph at $k_B T=1$ and for $b=1.72 \sigma$. 
The results are shown in Figure \ref{c_dens}. It is clear that $c$ decreases strongly with increasing density. Specifically, a $\sim 20\%$ density increase with respect to the melting density $\rho_{\rm melt}=0.479\sigma^{-3}$ leads to a vacancy concentration reduction of 4-5 orders of magnitude. As expected, an increasing density also leads to a slight reduction of the anharmonic effects.

\begin{figure} [c]

 \includegraphics[angle=-90,width=12.0cm]{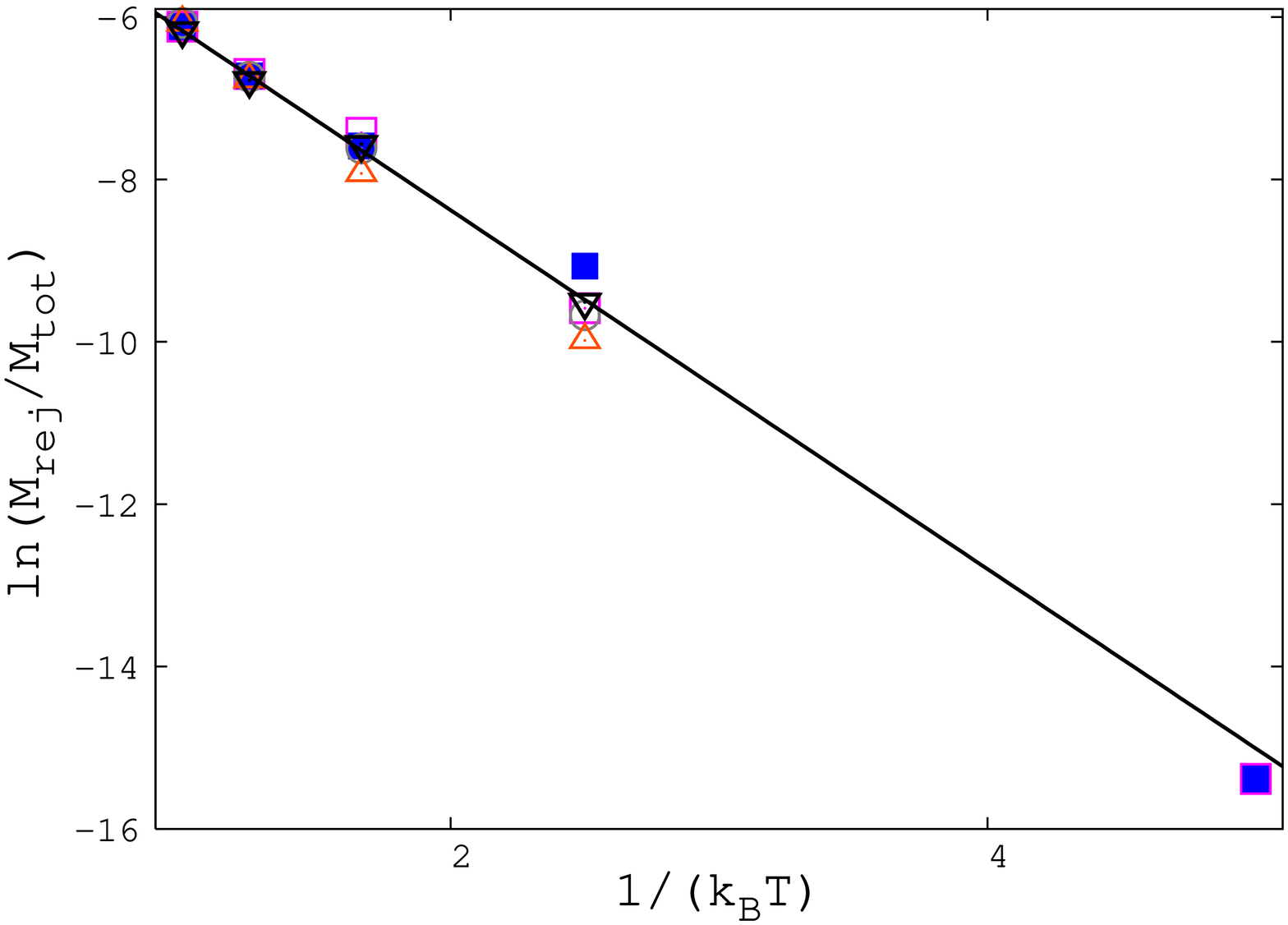}%

\caption{\label{wigseitz} Arrhenius plot of Wigner-Seitz rejection probability as a function of temperature. Results were obtained for $b=1.72\sigma$ and $N=108$ ({\color{magenta} $\square$}), $N=256$ ({\color{blue} $\blacksquare$}), $N=500$ ({\color{black} $\bigcirc$}), $N=864$ ({\color{Orange} $\bigtriangleup$}) and $N=1372$ ($\bigtriangledown$). The line represents the linear regression to the results obtained with $N=1372$.}

\end{figure}

As a final point, we have monitored the frequency of diffusion attempts of the twelve nearest-neighbors in the RW vacancy creation simulations. To this end, for $b=1.72 \sigma$, we recorded the number of times a trial move of any of the 12 nearest neighbors was rejected because of a Wigner-Seitz (WS) cell boundary crossing event as a function of temperature and the number of particles $N$ in the cell. The results are summarized in Fig.~\ref{wigseitz}, which presents an Arrhenius plot of the rejection probability $P_{WS}$ as a function of temperature. All the data fits very well to a straight line, irrespective of the number of atoms in the cell, allowing us to estimate the energy barrier $\Delta E$ that needs to be overcome for a vacancy to diffuse to a neighboring lattice site in the classical isomorph. Using $P_{\rm WS} \sim \exp(-\Delta E/k_B T)$, the slope of the Arrhenius plot gives $\Delta E=2.21$. While knowledge of the height of the energy barrier permits an estimation of the jump frequency in the classical isomorph, this evidently is not the case for the quantum system, for which the determination of tunneling frequencies requires knowledge of the shape of the potential-energy barrier.

\section{Conclusions}
\label{final}

In this paper we demonstrate the importance of anharmonic effects in calculating the zero-point vacancy concentration in solid $^{4}$He using the RW and HA methods based on the well-known quantum/classical isomorphism for the ground state of a many-body system of bosons. We find that anharmonicity leads to a decrease of the formation free energy of about $25\%$, resulting in a vacancy concentration that is more than an order of magnitude larger compared to the harmonic results.

We conclude by noting that the RW methodology for the calculation of free energies in classical isomorph is not restricted to the simple pseudopotential Jastrow form employed in this paper. In addition it can also be used to investigate the concentration of other species of defects such as interstitials and interactions between them. 

\section{Acknowledgments}
The authors gratefully acknowledge financial support from the Brazilian agencies FAPESP, CNPq and CAPES. Part of the computations were performed at the CENAPAD high-performance computing facility at Universidade Estadual de Campinas.


\begin{thebibliography}{31}
\bibitem[{Not()}]{Note}
\bibinfo{note}{In addition, we believe there is a typographical error in
  Ref.~\onlinecite{hod95}, in the which printed value of $b$ for the melting
  density does not correspond to the one given in Ref.~\onlinecite{hoo71}.}
\end{thebibliography}
\end{document}